\documentclass[reprint,twocolumn,notitlepage,prb,superscriptaddress,showpacs]{revtex4-2}

\usepackage[usenames,dvipsnames]{color}

\usepackage{comment}
\usepackage{graphicx}
\usepackage[utf8x]{inputenc}
\usepackage{lineno}
\usepackage{color}
\usepackage{hyperref}
\usepackage{amsmath}

\begin{document}

\title{One- and two-particle properties of the weakly interacting two-dimensional Hubbard model in proximity to the van Hove singularity}
\author{B. D. E. McNiven}
\affiliation{Department of Physics and Physical Oceanography, Memorial University of Newfoundland, St. John's, Newfoundland \& Labrador, Canada A1B 3X7} 
\author{Hanna Terletska}
\affiliation{Department of Physics and Astronomy, Computational Sciences Program,
Middle Tennessee State University, Murfreesboro, TN 37132, USA}
\author{G. T. Andrews}
\affiliation{Department of Physics and Physical Oceanography, Memorial University of Newfoundland, St. John's, Newfoundland \& Labrador, Canada A1B 3X7} 
\author{J. P. F. LeBlanc}
\email{jleblanc@mun.ca}
\affiliation{Department of Physics and Physical Oceanography, Memorial University of Newfoundland, St. John's, Newfoundland \& Labrador, Canada A1B 3X7}

\date{\today}
\begin{abstract}
We study the weak-coupling limit of the $t-t^\prime-U$ Hubbard model on a two-dimensional square lattice using a direct perturbative approach.  Aided by symbolic computational tools, we compute the longitudinal density-density correlation functions in the $\chi_{\uparrow \uparrow}$ and $\chi_{\uparrow \downarrow}$ basis from which we can obtain the dynamical spin and charge susceptibilities at arbitrary doping and temperature.
We find that for non-zero $t^\prime$, the zero frequency commensurate $\mathbf{q} = (\pi, \pi)$ spin and charge excitations are each strongest at different densities and we observe a clear behavioral change that appears tied to the van Hove singularity of the non-interacting dispersion upon which the perturbative expansion is built.  We find a strongly reduced compressibility in the vicinity of the van Hove singularity as well as a behavioral change in the double occupancy. For finite $t^\prime$, the observed van Hove singularity occurs away from half-filling leading us to conclude that that this reduction in compressibility is distinct from Mott insulating physics that one expects in the strong-coupling regime.  We compute the full dynamical spin and charge excitations and observe distinct structure for electron and hole doped scenarios in agreement with experiments on cuprate materials.  Finally, we observe a peculiar splitting in spin and charge excitations in the vicinity of the van Hove singularity, the origin of which is traced to a splitting near the bottom of the band.
\end{abstract}

\maketitle

\section{Introduction}
It is generally assumed that the dominant physics of the cuprate phase diagram emerges from strong electronic correlations, and for this reason, the square lattice Hubbard model in the strong-coupling regime has been the focus of theoretical and computational studies\cite{benchmarks,Huang2018,zheng:2017} as well as experimental work on ultra-cold atom systems.\cite{esslinger:2010, hulet:2020} Such work on the Hubbard model has found a plethora of phases with striking similarity to the doping dependent phase diagram of the high-temperature cuprate class of materials.\cite{gull:2013,zheng:2016,park:2008}
In contrast, much of the experimental work on cuprates connects single- and two-particle properties based on the arguments of nesting of scattering vectors for various Fermi surface topologies.\cite{vishik:2010}
If such nested momentum/energy transfer processes can be identified then the idea of vector nesting provides an incredibly powerful lens through which one can understand the effects of weak renormalization due to interactions\cite{nesting:1990}, and has often been a central consideration for the development of phenomenological theories of pseudogap physics.\cite{yrzreview,vishik:2018,pairdensity,leblanc:2011,leblanc:2014} 

Dynamical mean-field theory (DMFT)\cite{Georges} is perhaps the most well-known method for studying correlated electron systems which when applied to the 2D Hubbard model at strong coupling strengths, finds a gapped system known as a Mott insulator that originates from both strong local electron-electron interactions \emph{and}, in the case of the Hubbard model, proximity to the half-filling point of the band (see Supplemental Materials).  
A number of numerical studies have discovered that strong commensurate spin excitations lead to pseudogapped and insulating states in the symmetric ($t^\prime=0$) half-filled two-dimensional (2D) Hubbard model in the weak coupling limit, where direct perturbative schemes are convergent.\cite{Schaefer:2020,fedor:2020,behnam:2020,wu:2017}  
As temperature is decreased, the onset of insulating behavior occurs at smaller values of the Hubbard interaction strength, $U/t$, and also coincides with long-range antiferromagnetic spin correlation lengths.\cite{fedor:2021} This begs the question as to how one should interpret a wealth of literature that clearly demonstrates the formation of a Mott gap beyond the range of perturbative methods and above a finite critical interaction value,  $U_c$, in the $T=0$ limit.\cite{hubb:review:theory}  

For the single-band Hubbard model with only nearest-neighbour hopping, the strongest vector nesting is expected to occur at half-filling.  There is then the potential for a mixture of two physical phenomena occurring at the same point: insulating behavior due to vector nesting that is rooted in weak coupling ideas, and Mott-insulating physics rooted in the strong coupling limit. Separating these two effects is essential to developing a true understanding of the single-band model on the square lattice.

In order to delineate these two effects, we study the single-band model with finite next-neighbour hopping, $t^\prime$, which plays a key role since it moves the van Hove point for the non-interacting problem away from the half-filled point and breaks the perfect $\mathbf{q}=(\pi,\pi)$ nesting.  We then study the  charge and spin excitation spectra in the weak-coupling limit.  
Our methodology is based on the direct perturbative expansion of density-density correlation functions in the form of Feynman diagrammatics in conjunction with a scheme to automate the analytic evaluation of Matsubara sums.\cite{AMI,AMI:spin}   
This scheme, known as Algorithmic Matsubara Integration (AMI), provides analytic expressions in which Matsubara frequencies can be analytically continued by replacing $i\omega \to \omega+i\Gamma$  which is exact in the $\Gamma \to 0^+$ limit.  Further, the calculations do not suffer from finite-size effects, and the generated expressions have explicit dependence on temperature and chemical potential, ultimately giving access to arbitrary temperatures and doping which when combined, allow us to develop a full picture of the weak-coupling physics within the model.

\section{Methods}

\subsection{Hubbard Hamiltonian}
We study the single-band Hubbard Hamiltonian on a 2D square lattice\cite{benchmarks},
\begin{eqnarray}\label{E:Hubbard}
H = \sum_{ ij \sigma} t_{ij}c_{i\sigma}^\dagger c_{j\sigma} + U\sum_{i} n_{i\uparrow} n_{i\downarrow},
\end{eqnarray}
where $t_{ij}$ is the hopping amplitude, $c_{i\sigma}^{(\dagger)}$ ($c_{i\sigma}$) is the creation (annihilation) operator at site $i$, $\sigma \in \{\uparrow,\downarrow\}$ is the spin, $U$ is the onsite Hubbard interaction, $n_{i\sigma} = c_{i\sigma}^{\dagger}c_{i\sigma}$ is the number operator.  We restrict the sum over sites to nearest and next-nearest neighbors for a 2D square lattice, resulting in the free particle energy 
\begin{eqnarray}
\nonumber\epsilon(\textbf k)=-2t[\cos(k_x)+\cos(k_y)]-4t^\prime [\cos(k_x)\cos(k_y)]-\mu,
\end{eqnarray} 
where $\mu$ is the chemical potential, and $t$($t^\prime$) is the nearest (next-nearest) neighbor hopping amplitude.  Throughout, we work with energies in units of the hopping, $t=1$. 

\subsubsection{van Hove Singularity}
The primary effect of a non-zero $t^\prime$ is that the van Hove singularity will occur at a density away from half-filling.  For the non-interacting case this can be found analytically from $\epsilon(\mathbf{k})$.  For values of $|t^\prime|<0.5$, the van Hove singularity occurs at a chemical potential of $\mu=4t^\prime$.  For larger amplitudes of $t^\prime$, the topology of the Fermi surface is changed substantially from the $t^\prime=0$ case, because the next-nearest neighbor hopping becomes dominant.\cite{Romer15}  We therefore restrict our study to nominal values of $|t^\prime|<0.5$.

In the case of non-interacting problems there is a one-to-one correspondence between the van Hove singularity and the location of a topological change in the Fermi surface known as a Lifshitz transition.  Past studies of the Lifshitz transition in the 2D Hubbard model have focused on the large $U/t$ insulating regime, where even for $t^\prime=0$, the Lifshitz transition occurs for densities $\langle n\rangle \equiv n <0.5$.\cite{jarrell:lifshitz,norman:2010} 
Our calculations represent a perturbative expansion built upon the Hartree-shifted, but otherwise non-interacting, problem.  The inclusion of a Hartree-shift creates a $U/t$ dependence in the relationship between chemical potential, $\mu$, and density, $n$. 
Since we are at the weakly interacting limit of the model, the system has a metallic Fermi surface, and therefore the location of the van Hove point is not changed substantially from the non-interacting case, and should always occur in the vicinity of $\mu=4t^\prime$.  We therefore use this information to guide our choice of parameters throughout.

\subsection{Perturbation Expansion} We obtain the diagrams  for the spin susceptibility $\chi_s=\langle \mathcal{T} \hat{S}_z(\tau, x)\hat{S}_z(\tau^\prime, x^\prime) \rangle$ and for the charge susceptibility $\chi_d=\langle \mathcal{T} \hat{n}(\tau, x)\hat{n}(\tau^\prime, x^\prime) \rangle$ via perturbative expansions of each set of operators.  The two expansions are related due to the definitions of $\hat{S}_z=\hat{n}_\uparrow - \hat{n}_\downarrow$ and $\hat{n}=n_\uparrow+ n_\downarrow$.  We can therefore define the susceptibility in a basis of correlations between $\hat{n}_\uparrow$ and $\hat{n}_\downarrow$ operators. Assuming spin symmetry, we denote relevant correlation functions as  $\chi_{\uparrow \uparrow}=2\langle  \mathcal{T} \hat{n}_\uparrow(\tau,x) \hat{n}_\uparrow(\tau^\prime, x^\prime)\rangle$ and $\chi_{\uparrow \downarrow}=2\langle \mathcal{T} \hat{n}_\uparrow(\tau,x) \hat{n}_\downarrow(\tau^\prime, x^\prime)\rangle$.\cite{Rohringer12}  
This leads to the simple relations $\chi_s=\chi_{\uparrow \uparrow} - \chi_{\uparrow \downarrow}$ and $\chi_d=\chi_{\uparrow \uparrow} + \chi_{\uparrow \downarrow}$. 

We present as well the double occupancy $D=\langle n \rangle^2+2\langle \hat{n}_\uparrow(\tau, x)\hat{n}_\downarrow(\tau, x) \rangle$.\cite{kung:2015}  The first term 
represents the uncorrelated disconnected diagrams while the second term is the local, same-time contribution from interactions.  The second term can be obtained from  
$\sum\limits_{q} \sum\limits_{n}\chi_{\uparrow \downarrow}(\mathbf{q},i\Omega_n)$. We provide a summary of each diagrammatic expansion in the supplemental materials.  The convergence rate for each observable can differ drastically.  For example, results for double occupancy and density as well as quantities on the Matsubara axis are much easier to compute while results on the real frequency axis take substantially more computational effort.  Results are 
obtained to fourth order in the interaction for static and Matsubara axis properties, and to third order for real-frequency 
evaluation. We note that there is no conceptual hurdle associated with extending to higher orders, but there is a computational hurdle that is beyond exponential in the expansion order.  Further advancements will be required to overcome those hurdles, such as those suggested in Refs.~\onlinecite{igor:spectral} and \onlinecite{jaksa:analytic}.

\subsection{Algorithmic Matsubara Integration} First presented in Ref.\cite{AMI}, AMI automates the evaluation of internal Matsubara sums for arbitrary Feynman diagrams via a repeated application of the well-understood residue theorem.  This works so long as the perturbative expansion can be built within a diagonal basis of known eigenvalues that are frequency independent.  The interaction must also be frequency independent, or its frequency dependence must be explicitly known. 
In the case of the Hubbard interaction, the result of AMI applied to an $n$th order diagram is an analytic expression comprised of: a prefactor times $U^n$; a product of Fermi/Bose distribution functions and derivatives of such; and a product of non-interacting Green's functions.  Each diagram typically results in many such terms, the number of which grow exponentially with expansion order (see Supplemental Materials).  
We use AMI to analytically perform the Matsubara sums over the internal Matsubara frequencies for $\chi_{\uparrow \uparrow}$ and $\chi_{\uparrow \downarrow}$, and also the external frequency $i\Omega_n$ for the double occupancy. The resulting analytic expressions must then be integrated numerically over the remaining internal spatial degrees of freedom, which we resolve using standard Monte-Carlo techniques for which we make use of the open source \texttt{ALPSCore} framework\cite{gaenko:2017,alpscore_v2} combined with the open source AMI library \texttt{libami}.\cite{libami}

\subsubsection{Analytic Wick Rotation to the Real Frequency Axis.} The results of AMI are analytic expressions containing the external frequency that can be analytically continued to the real frequency axis without numerical methods such as maximum entropy inversion \cite{maxent,jarrell:maxent}.   In the case of frequency dependent observables, we perform analytic continuation by replacing the external frequency $i\Omega_n \to \omega +i\Gamma$ which is exact in the $\Gamma \to 0^+$ limit. Throughout, we employ a finite value of $\Gamma=0.125$ which serves as a numerical regulator.  The impact of the regulator can be controlled, with larger values of $\Gamma$ acting to soften sharp features.  The regulator should typically appear as the smallest energy scale to ensure it does not impact results.

\subsubsection{Numerical Analytic Continuation to the Real Frequency Axis.} While numerical analytic continuation is not required in our approach, we can equally well produce results on the Matsubara axis, and perform numerical analytic continuation.  For this we employ the method of maximum entropy inversion using a flat default model\cite{jarrell:maxent} utilizing the code presented in Ref.~\onlinecite{maxent} and its dependencies.\cite{gaenko:2017,alpscore_v2}

\begin{figure}
    \centering
    \includegraphics[width=\linewidth]{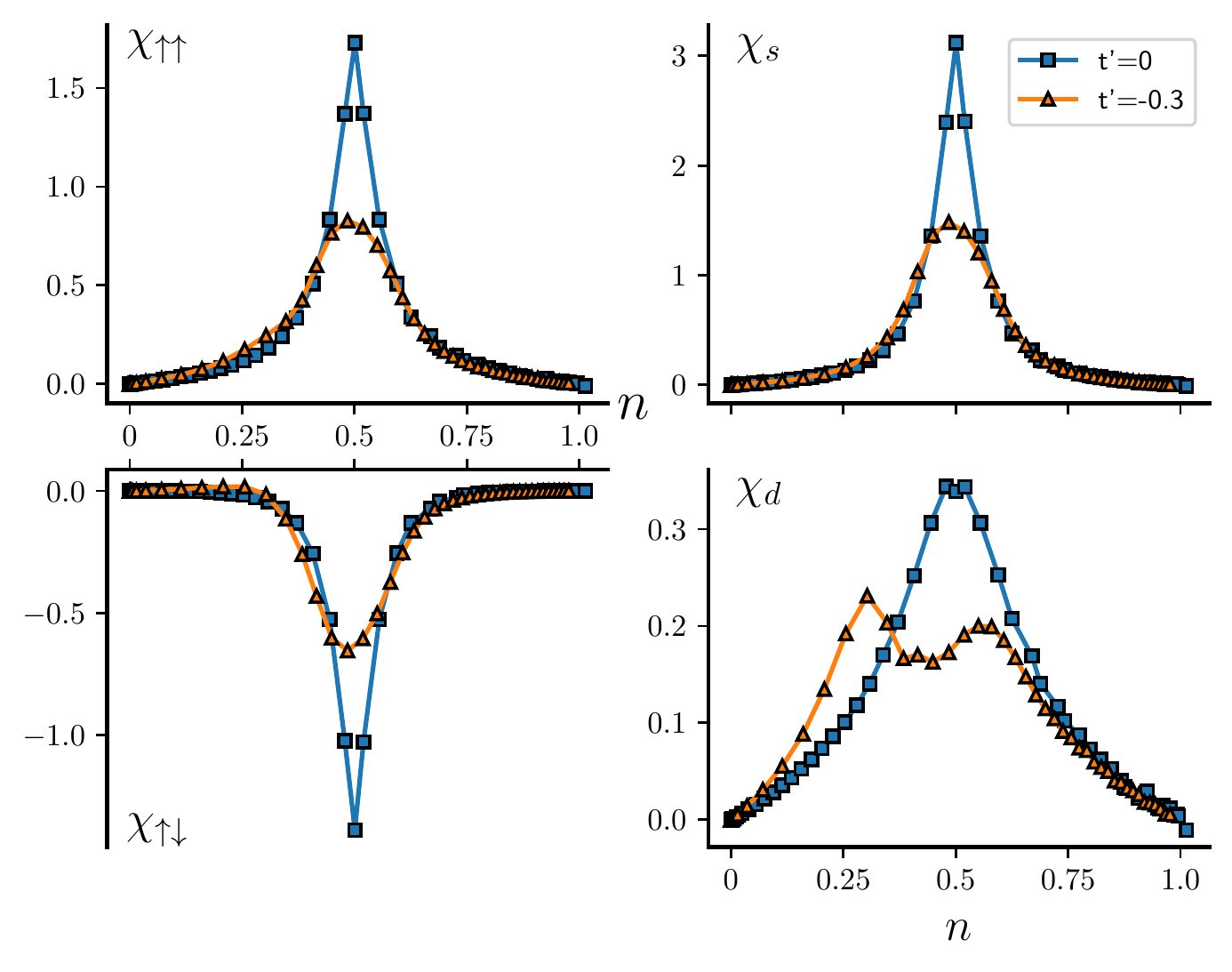}
    \caption{\label{fig:fig1p1} The static two-particle susceptibility at scattering vector $\mathbf{q}=(\pi,\pi)$ in the $\chi_{\uparrow \uparrow}$, and $\chi_{\uparrow \downarrow}$ basis (left) and in the spin/charge basis (right) for $t^\prime=0.0, -0.3t$, $U/t=3$, $\beta t=5$ as a function of particle density, $n$.  }
\end{figure}


\smallskip
\section{Results}
\subsection{Doping and Temperature Dependence of Static $\mathbf{q}=(\pi,\pi)$ Susceptibilities.}  We present in Fig.~\ref{fig:fig1p1} the density dependence of the static $\mathbf{q} = (\pi, \pi)$ susceptibility in the $\uparrow\uparrow/\uparrow\downarrow$, as well as the spin/charge bases (annotated by $s/d$ following Ref.~\cite{Rohringer12}) for the particle-hole symmetric case of $t^\prime=0$, and for a particle-hole asymmetric case of $t^\prime=-0.3t$, both at $\beta t=5$ for a nominal interaction strength of $U/t=3$.  Considering first $\chi_{\uparrow\uparrow}$ and $\chi_{\uparrow\downarrow}$, we see that the former is positive for all densities and peaked at the half-filling point $n=0.5$ while the latter is negative for all densities.  The simple  subtraction or addition of these two curves leads directly to the spin and charge susceptibilities\cite{Rohringer12}, respectively,  shown in the right hand frame of Fig.~\ref{fig:fig1p1}.  For $t^\prime=0$, the perfect particle-hole symmetry results in both the spin and charge susceptibility being strongest at half-filling.  This is not the case for finite $t^\prime$ where we see that moderate asymmetry leads to a mismatch between $\chi_{\uparrow \uparrow}$ and $\chi_{\uparrow \downarrow}$ peaks.  When obtaining $\chi_s$ this slight asymmetry is washed out, and while it does decrease with $t^\prime$, the spin susceptibility remains peaked near the half-filled point.  This is not the case for the charge susceptibility, $\chi_d$, where we see that the simple addition of $\chi_{\uparrow\uparrow}$ and $\chi_{\uparrow\downarrow}$ leads to a dip in susceptibility near half-filling and creates a structure with one peak on the electron-doped side and a second peak on the hole-doped side.  We have performed additional calculations from non-perturbative methods to verify the existence of the two-peak structure of $\chi_d$ (See Supplemental Materials).

\begin{figure}
    \centering
    \includegraphics[width=\linewidth]{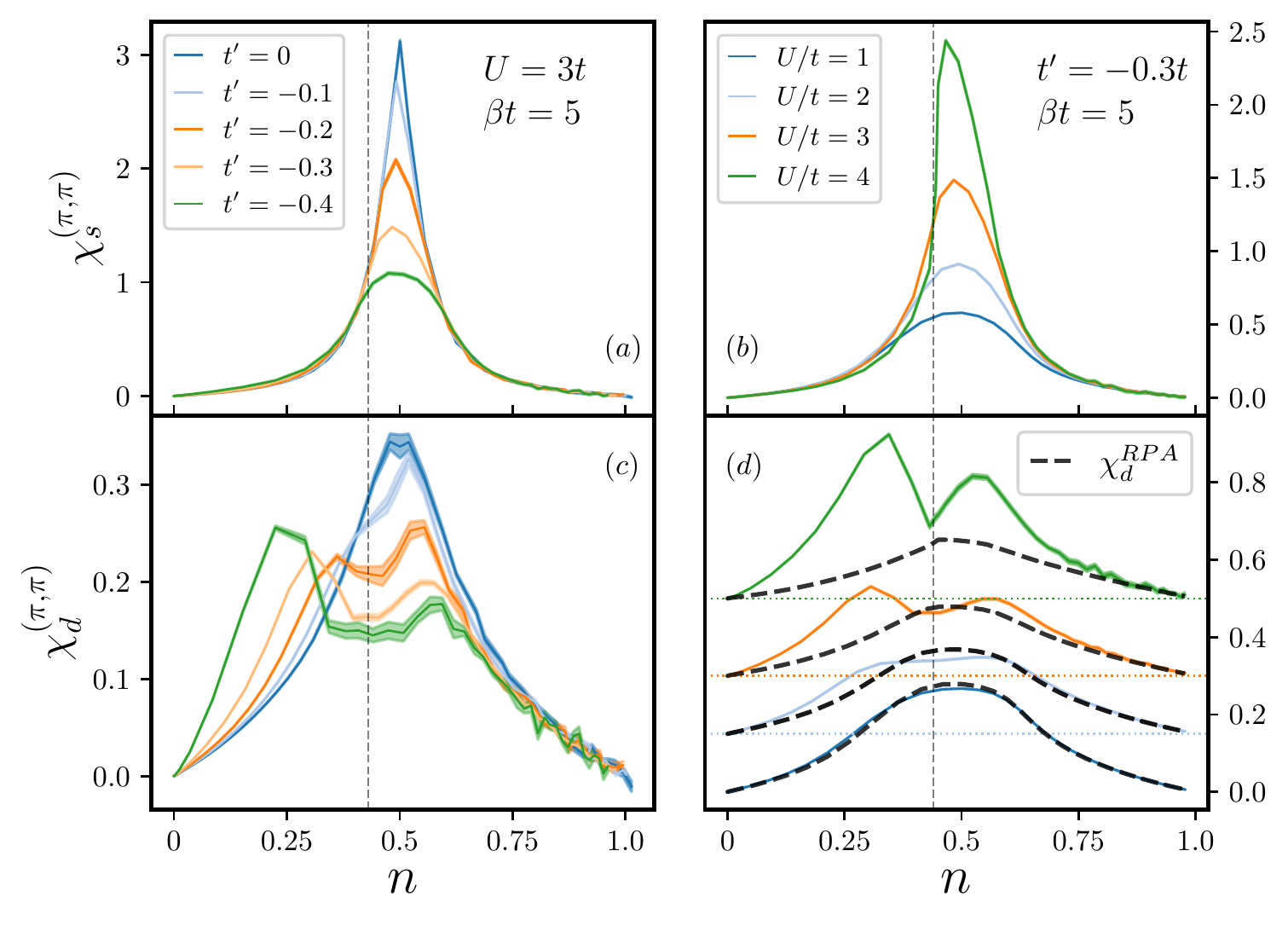}
    \caption{\label{fig:betaU}The static spin (top row) and charge (bottom row) susceptibilities at $\mathbf{q}=(\pi,\pi)$, $t^\prime=-0.3t$ as functions of doping for: $(a/c)$  fixed $U/t=3$ for variation in $t^\prime$; and $(b/d)$ for fixed $t^\prime$  for variation in $U/t$.  Vertical lines represent the location of the van Hove singularity for $t^\prime=-0.3$ at $U/t=3$ in the case of frames $(a/c)$ and for $U/t=4$ in the case of frame $(d)$ .  Black dashed curves in $(d)$ are RPA results for the corresponding interaction strengths.  $\chi_d$ and $\chi_d^{RPA}$ curves in frame (d) are offset for clarity.  }
\end{figure}

Precisely how the two-peak structure in $\chi_d$ is shaped is dependent upon the degree of asymmetry (value of $t^\prime$) but also on the temperature and interaction strength.  We present results for $\chi_s^{(\pi,\pi)}$ and $\chi_d^{(\pi,\pi)}$ in Fig.~\ref{fig:betaU} for variation in $t^\prime$ at fixed interaction strength of $U=3t$ (left) and at fixed temperature ($\beta t = 5$) for variation in interaction strength (right).  
Beginning with commensurate spin-excitations in Fig.~\ref{fig:betaU}(a), we see the dominant spin peak at $n=0.5$ for the most widely studied case of $t^\prime=0$.  Increasing the magnitude value of $t^\prime$ incrementally causes a reduction in $\chi_s^{(\pi,\pi)}$ in the vicinity of $n=0.5$ and has virtually no effect below $n=0.4$ or above $n=0.6$.  In contrast, we see an entirely different behavior from $\chi_d^{(\pi,\pi)}$ shown in Fig.~\ref{fig:betaU}(c).  While the peak in commensurate charge excitations for the $t^\prime=0$ case occurs at $n=0.5$, even a modest change in $t^\prime$ causes a depletion of charge excitations which results in a two-peak structure as a function of doping.\cite{gull:2020:charge}  Interestingly, the depletion is not centered at the half-filled point, but instead in the vicinity of the van Hove singularity marked with a vertical dashed line. 
Shown in Fig.\ref{fig:betaU}(d), this splitting appears to be a robust feature that exists above $U=2t$ for increasing interaction strengths.  Similarly, there is no structure in the spin susceptibility.
For reference, we also include results for the charge susceptibility from the simplistic random-phase approximation, $\chi_d^{RPA}$, marked as black dashed curves in Fig.~\ref{fig:betaU}~(d).  At weak $U/t=1$ the AMI and RPA results are nearly identical, but for larger $U/t$ the RPA result does not demonstrate a two-peak structure.  While the RPA expansion does include some diagrams from $\chi_{\uparrow \downarrow}$ it does so only through even powers of the bare bubble diagram.  This results in no mismatch in peaks of $\chi_{\uparrow \uparrow}$ and $\chi_{\uparrow \downarrow}$ as described in Fig.~\ref{fig:fig1p1} and this results in only a single peak in the density dependence.



It is perhaps not surprising that we observe the maximal $\mathbf{q}=(\pi,\pi)$ charge excitations, at low temperatures, near the van Hove singularity since there the Fermi surface will most closely resemble that of the antiferromagnetic Brillouin zone and allow for vector nesting in the susceptibility.
There is evidence to suggest that insulating behaviour in the 2D Hubbard model at weak coupling is caused by strong $\mathbf{q}=(\pi,\pi)$ antiferromagnetic fluctuations\cite{wu:2017,behnam:2020,gunnarsson:2015,Schaefer:2020}, 
which in Fig.~\ref{fig:betaU} remain centered at half-filling.  If pseudogap and insulating behavior are indeed due to antiferromagnetic fluctuations then we expect to find insulating behavior near half-filling.  We will see in the next section that this assertion appears to be false, and that the suppression of the charge susceptibility, that is an indicator of insulating character, is centered at the van Hove point, and coincides with a reduced compressibility of the electron density.

\subsection{Double Occupancy and Compressibility}   
\begin{figure}
    \centering
    \includegraphics[width=\linewidth]{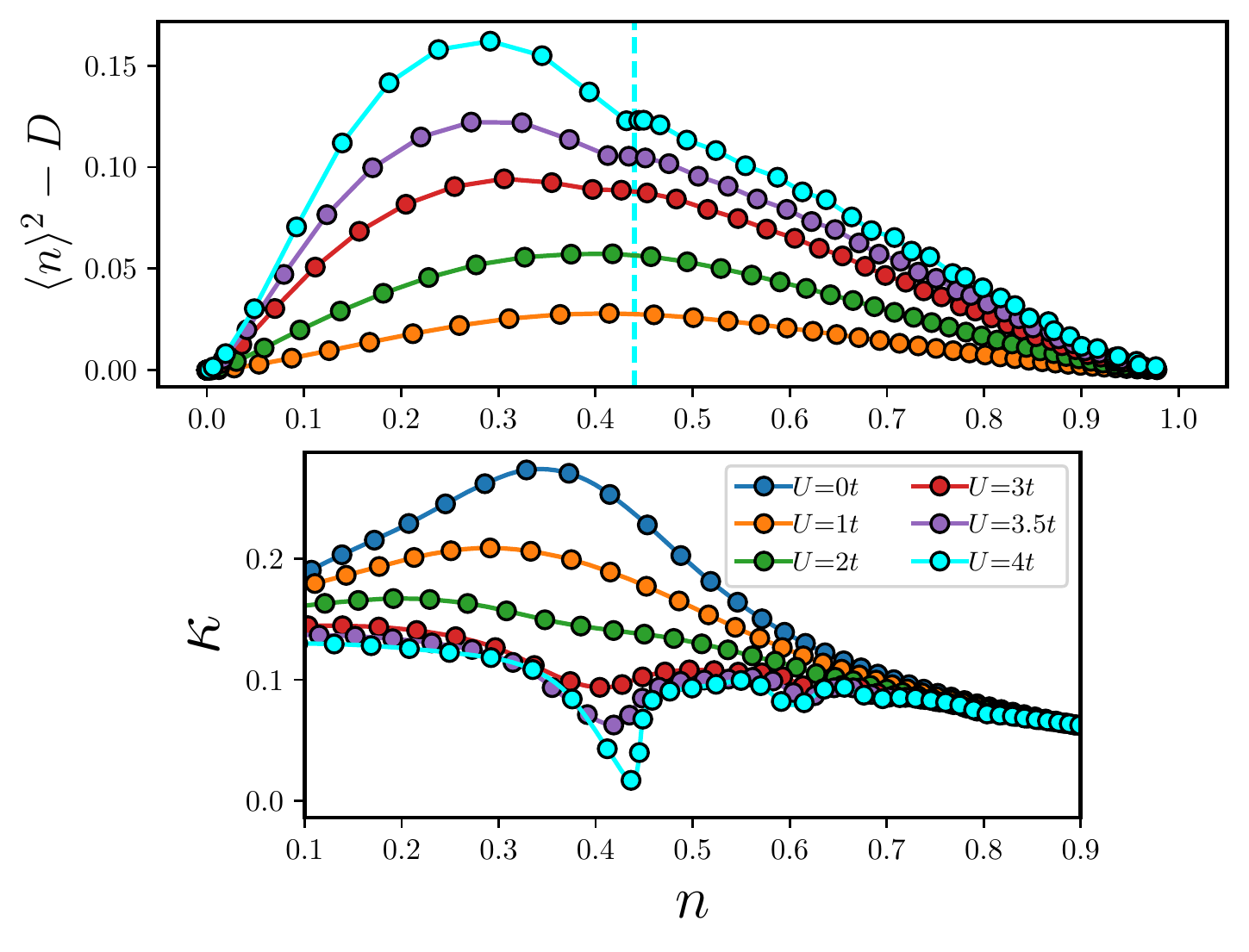}
    \caption{  \label{fig:Dkappa}Deviation of double occupancy from the uncorrelated case (top) and compressibility (bottom)  as a function of density for fixed $t^\prime=-0.3t$,  $\beta t=5$.  Vertical dashed line marks location of van Hove singularity at $U/t=4$. }
\end{figure}

Representing susceptibilities in the $\uparrow \uparrow$/$\uparrow \downarrow$ basis is particularly useful since it separates out the key element that distinguishes the charge and spin susceptibilities, namely contributions from the correlated  $2\chi_{\uparrow \downarrow}$.  One can get an imprint of the impact of $\chi_{\uparrow \downarrow}(\mathbf{q},\Omega)$ by examining also its same-time, local counterpart the double occupancy, $D$.  
For a non-interacting system, one would find $\chi_{\uparrow\downarrow}=0$ and the double occupancy is given by $D=\langle n\rangle ^2$.  For the interacting system, it is convenient to consider the deviation from the non-interacting case $\langle n\rangle ^2-D$.\cite{kung:2015}
We present this deviation in Fig.~\ref{fig:Dkappa}~(top) for the particle-hole asymmetric case for $t^\prime=-0.3t$ at nominal temperature of $\beta t=5$ for increasing interaction strength.  
Here, a positive value of the deviation corresponds to a reduced value of the double occupancy which we see occurs for all densities, similar to the observed negative value of $\chi_{\uparrow\downarrow}$ in Fig.~\ref{fig:fig1p1}.  One expects to find a maximal reduction in double occupancy occurring when interaction effects are maximal.  Above $U=2t$, the data exhibits a peak near $n=0.3$ and demonstrates a kink feature at slightly higher density. This kink feature coincides with the van Hove  singularity which for $U=4t$ occurs at a density of $n\approx0.44$ (marked with vertical dashed line in Fig.~\ref{fig:Dkappa}).

\begin{figure*}
    \centering
    \includegraphics[width=0.8\linewidth]{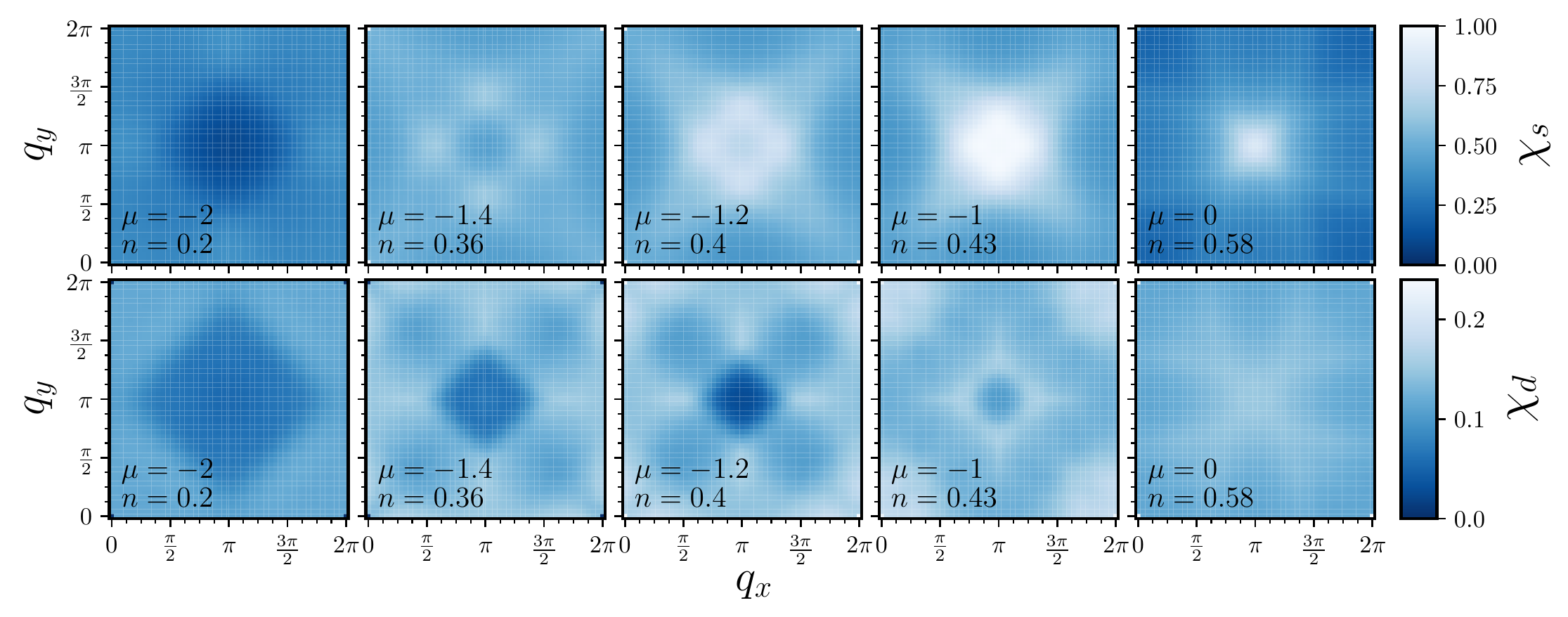}\\
      \caption{\label{fig:chi_panel} $\chi_s(\Omega=0,\mathbf{q})$ (top) and $\chi_d(\Omega=0,\mathbf{q})$ (bottom) as a function of $q_x$ and $q_y$  for $U/t=2.5$, $t^\prime=-0.3t$, and $\beta t=5$ for a range of densities $n=0.2\to 0.58$. }
 \end{figure*}

We present also the compressibility, $\kappa=\frac{\partial n}{\partial \mu}$, as a function of density in the lower frame of Fig.~\ref{fig:Dkappa}.  Upon increasing interaction strength, we observe a reduction in compressibility that produces a minimum near the van Hove point.  This reduction in compressibility grows stronger for increasing value of $U/t$ and leads ultimately to an incompressible phase centered precisely at the van Hove point.  We mark the van Hove point for $U/t=4$ in the top frame of Fig.~\ref{fig:Dkappa} and see that it coincides with a minima in the charge excitations displayed in the right hand frames of Fig.~\ref{fig:betaU} and not with the maxima of spin excitations.  In the limit of strong coupling we expect that the role of spin excitations will dominate and this feature would migrate to the half-filled point as seen from non-perturbative methods (see Supplemental), and similarly as one removes the particle-hole asymmetry due to $t^\prime$.

Reiterating our earlier remark, the observation of incompressibility at the van Hove point and not at half-filling, where spin excitations are strongest, conflicts with the assertion that it is $\mathbf{q}=(\pi,\pi)$ spin fluctuations that are responsible for pseudogap and insulating behavior seen in Refs.~\cite{Schaefer:2020} and \cite{fedor:2020}.  Past works studied primarily the $t^\prime=0$ case where the van Hove and half-filled points coincide.  Here with finite $t^\prime$ we see that there are two distinct mechanisms causing insulating behavior, that of strong spin excitations and the role they are expected to play in pseudogap physics, and that of a peaked density of states that occurs near the van Hove singularity of the non-interacting system.


\subsection{Momentum Dependence of Static Spin and Charge Excitations}

\begin{figure*}
    \centering
    \includegraphics[width=0.8\linewidth]{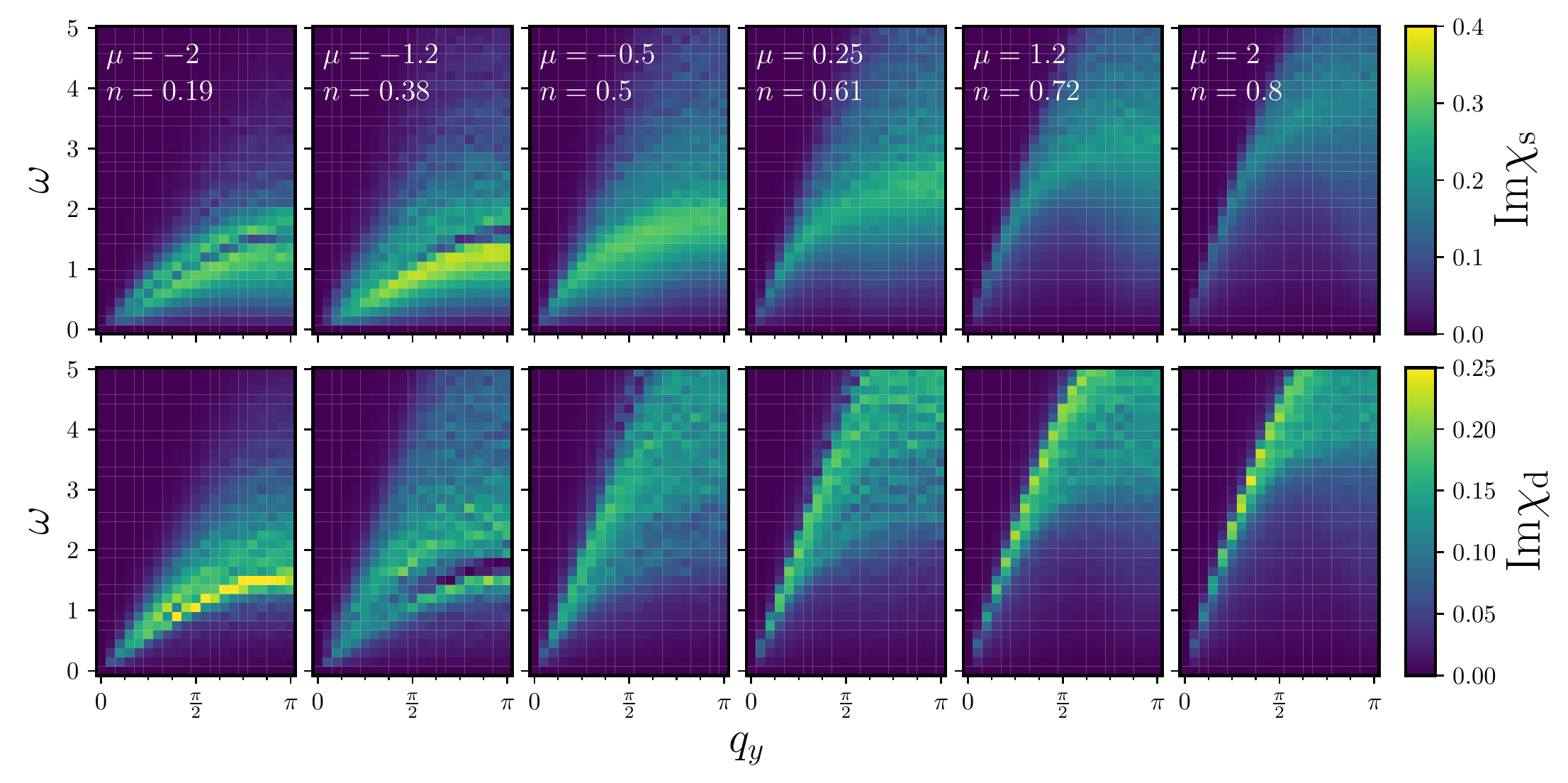}\\
      \caption{\label{fig:realfreq}Imaginary parts of $\chi_s(\omega,q)$ (top) and $\chi_d(\omega,q)$ (bottom) as a function of $\omega$ and $\mathbf{q}=(0,q_y)$  for $U/t=2.5$, $t^\prime=-0.3t$, and $\beta t=5$ for a range of densities $n=0.19\to 0.8$.  }
 \end{figure*}

\begin{figure}
    \centering
      \includegraphics[width=0.9\linewidth]{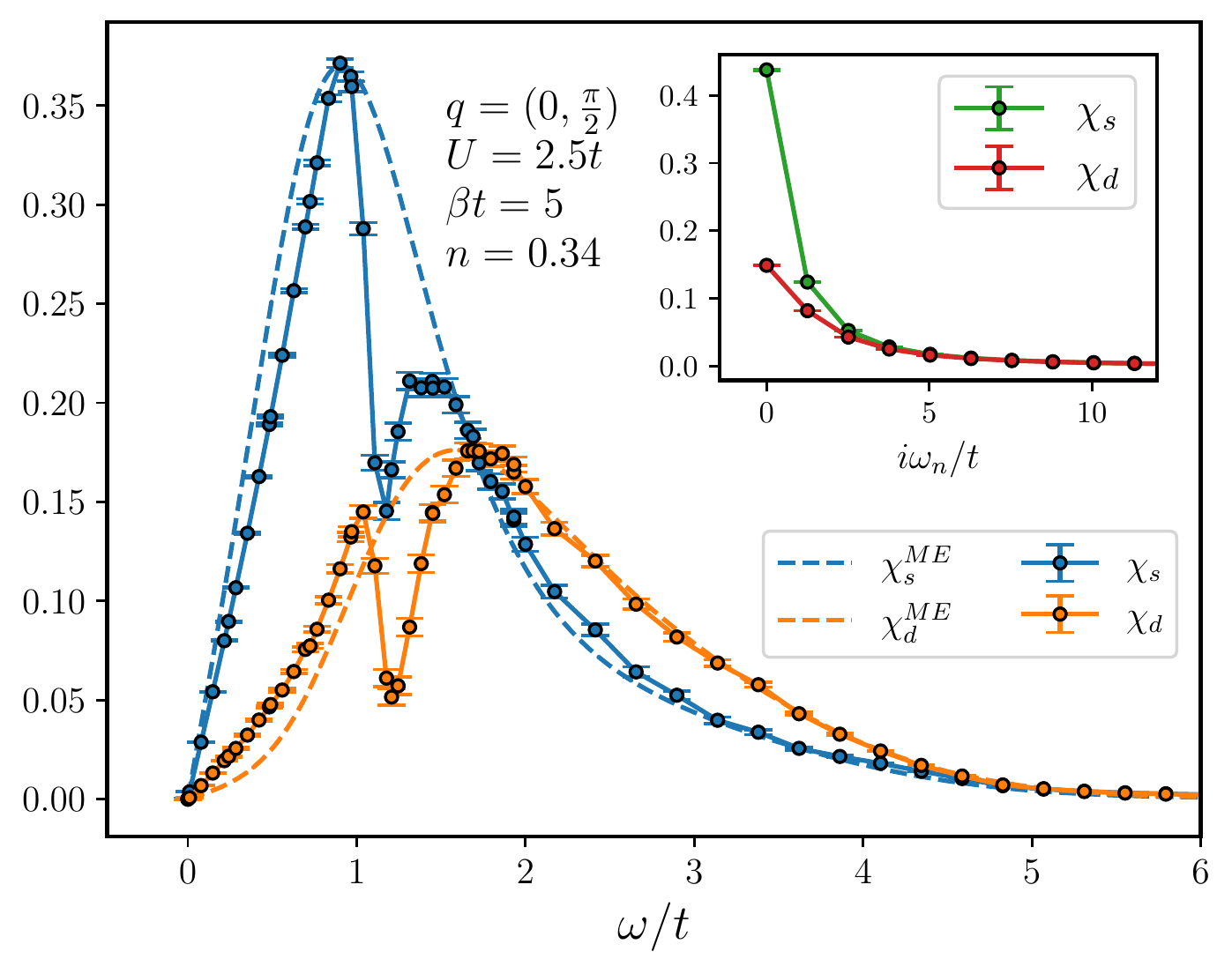}
    \caption{ \label{fig:realcut} The imaginary parts of $\chi_s(\omega,q)$ and $\chi_d(\omega,q)$ at $\mathbf{q}=(0,\pi/2)$ for a density of $n=0.34$ produced via analytic continuation of $i\omega_n\to \omega+i\Gamma$ for the case of $\Gamma=0.125$.  The result of numerical analytic continuation via maximum entropy inversion ($ME$) is shown for reference with input from $\chi_s(i\omega_n)$ and $\chi_d(i\omega_n)$ shown in the inset.}
\end{figure}

We present in Fig.~\ref{fig:chi_panel} results of the static spin (top row) and charge (bottom row) susceptibility for the $U/t=2.5$ case, plotted as a function of the scattering vector $\mathbf{q}$.  This choice of interaction strength is large enough to show distinct spin and charge behavior while keeping numerical uncertainty under control.  For densities above half filling the spin and charge susceptibilities appear similar, with both having maximal values at $\mathbf{q}=(\pi,\pi)$.  The spin susceptibility at $(\pi,\pi)$, however, is substantially larger than at other $q$-vectors while the charge susceptibility is much more diffuse.  As one reduces chemical potential towards the van Hove point at $\mu=-1.2$, we primarily note the separation of the single $\mathbf{q}=(\pi,\pi)$ spin excitation into two distinct peaks at $\mathbf{q}=(\pi\pm \delta,\pi)$ and $\mathbf{q}=(\pi,\pi\pm \delta)$ as noted in previous works \cite{behnam:2020,fedor:2021}.  Reducing $\mu$ further causes $\delta$ to increase and this moves the peaks further from the commensurate $(\pi,\pi)$ point. One also observes secondary structure in the spin susceptibility that is strongest along the diagonals - qualitatively similar to observations in past RPA studies.\cite{Romer15}
A similar, albeit weaker, splitting near $\mathbf{q}=(\pi,\pi)$ is observed in $\chi_d$.  Rather than distinct peaks, we observe a nearly continuous diamond shape and primarily demonstrates an increase in signal along the boundaries, particularly near $\mathbf{q}=(0,0)$.

A sharp (in momentum) peak should be observed in the case of long-ranged but static correlations.  In our data,  the  absence of strong features in neither the charge nor the spin  susceptibility below the van Hove point suggest that the origin of incompressibility seen in Fig.~\ref{fig:Dkappa} is not a property of static nesting.  This leaves only then the possibility of dynamical scattering which we explore in Sec.~\ref{sec:dynamic}.

We note as well that the values of $\chi_s$ and $\chi_d$ at the $\Gamma$ point of the Brillouin zone are not zero.  This is due to the order of the limits $\lim_{q\to0}\lim_{\Omega\to0}$, where for static susceptibilities the zero frequency limit is 
applied before the zero scattering-vector limit. Our results for dynamic susceptibilities ($\Omega \neq 0$) correctly represent the reverse order of the limit where for all non-zero frequencies the susceptibilities are zero when $\mathbf{q}=(0,0)$.


\smallskip 
\subsection{Dynamical Spin and Charge Collective Excitations}\label{sec:dynamic}  Within our approach, the same calculations that provide the static $\mathbf{q} = (\pi, \pi)$ susceptibilities for arbitrary densities can be used to produce both finite real-frequency and Matsubara frequency results, in the thermodynamic limit, at arbitrary $q$-vectors.  
Access to real frequency susceptibilities allows us to examine the dispersive behaviour of plasmon and magnon collective excitations.  Much is known of charge excitations from studies of the two-dimensional electron gas\cite{igor:spectral} where the RPA chain of diagrams has a formal divergence resulting in a sharp quasiparticle peak.  In this work we operate with a truncated expansion and therefore interpret peaks in the imaginary part of the charge susceptibility as plasmon excitations and similarly peaks in the spin susceptibility as magnon excitations.  This interpretation is commonplace in the case of spin excitations where the cross-section for magnetic neutron scattering  is related to the imaginary part of the spin susceptibility.\cite{fong:1996}  

We plot the dispersions of the spin and charge excitations in Fig.~\ref{fig:realfreq} for a range of doping at $U/t=2.5$ and $\beta t=5$ for momenta along the $\mathbf{q}=(0,q_y)$  direction.
Beginning with the strongly electron doped case at $n=0.8$, when accounting for the change in scale the dispersion of $\chi_s$ and $\chi_d$, are very similar, with both showing a linear form up to rather high energies.  Since the difference between $\chi_s$ and $\chi_d$ is just $2\chi_{\uparrow\downarrow}$, this similarity suggests that $\chi_{\uparrow \uparrow}$ is the dominant contribution far from half-filling.  Reducing the density towards half-filling ($n=0.5$), the charge susceptibility remains linear but shows more incoherent signal at lower energies.  This incoherence coincides with peaks in $\chi_s$ that show a more recognizable sine-function shape that is representative of linear spin-wave models.\cite{coldea:2001,LeBlanc:2019,macdonald:1988}
Moving to the hole-doped cases, we observe a massive behavioral change at $n=0.38$ (coinciding with the van Hove point at $\mu=-1.2$), where both the spin and charge excitations have split into two bands at finite $q$-vectors, and merge at small $q$-vectors and low energy. The observation of splitting in $\chi_s$ and $\chi_d$ near the van Hove point suggests that this effect is likely due to the incompressibility observed in Fig.~\ref{fig:Dkappa} where $\kappa\to0$. This impacts the susceptibility due to renormalization of the single-particle propagator which occurs in diagrams that are part of the $\chi_{\uparrow \uparrow}$ expansion,  the amplitude of which is shared by $\chi_s$ and $\chi_d$. The splitting is therefore not due to vertex effects that are primarily a part of the $\chi_{\uparrow \downarrow}$ expansion (see Supplemental Materials).
Reducing the density further to the dilute limit, $n=0.19$, we see that the two-peak structure has closed in $\chi_d$ but remains in $\chi_s$.  We note a similarity in the overall shape of the dispersions of $\chi_s$ and $\chi_d$ as seen in the heavily electron doped case, suggesting again that $\chi_{\uparrow\uparrow}$ is the dominant contribution and that vertex effects are minimal.

\begin{figure}
    \centering
      \includegraphics[width=0.9\linewidth]{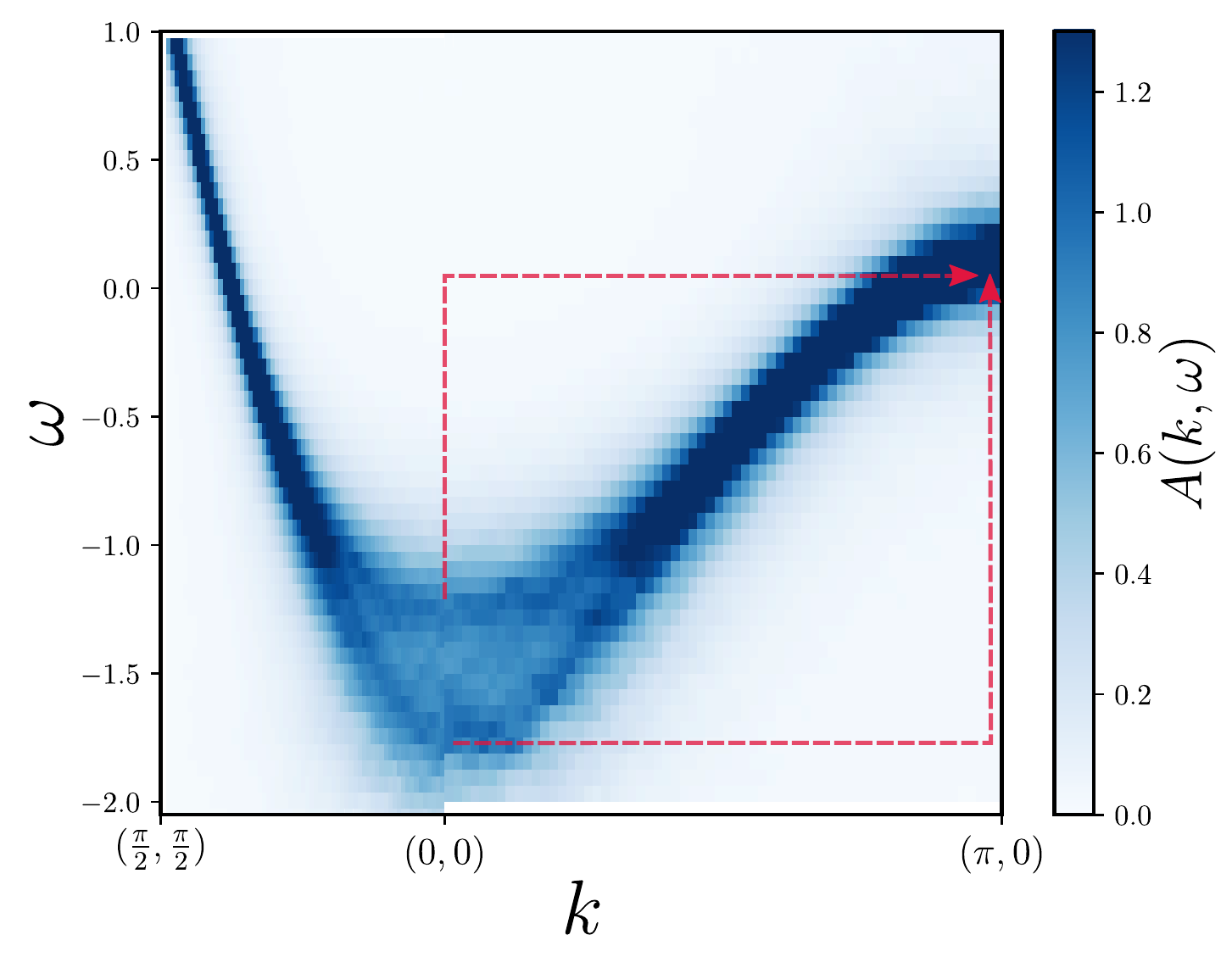}
    \caption{ \label{fig:spectralfunction}The spectral density, $A(k,\omega)$, along the path $\mathbf{k}=(\pi/2,\pi/2)\to(0,0)\to X$ for the case shown in Fig.~\ref{fig:realcut}.  Dominant non-zero frequency transitions from the bottom of the band to the $\mathbf{k}=(\pi,0)$ anti-nodal point are marked in red. }
\end{figure}

We have tracked this splitting with density and find that it is strongest in proximity to the van Hove singularity that coincides with the associated incompressible phase shown in Fig.~\ref{fig:Dkappa}.  We present a representative case in Fig.~\ref{fig:realcut} (see Supplemental Materials for additional data), where a frequency cut of the dispersion at fixed $\mathbf{q} = (0, \pi/2)$ for a hole-doped case with $\langle n \rangle =0.4$ is shown for an interaction strength of $U/t=2.5$ at $\beta t =5$.  We plot $\rm{Im}\chi_{s/d}(\omega+i\Gamma)$ in the main frame but also the result on the Matsubara axis, $\rm{Re}\chi_{s/d}(i\omega_n)$, in the inset. We stress that real-frequency and Matsubara axis results are the evaluation of the same analytic expressions.  For connection to earlier works we present also the numerical analytic continuation of $\chi_{s/d}(i\omega_n)\to \chi_{s/d}^{ME}(\omega)$ via maximum entropy (ME) inversion.  Results from ME are reminiscent of previous studies of the model\cite{kung:2015,jia:2014,huang:2017} and while we see that the numerical analytic continuation has the same general shape and dominant peak location as the direct real-frequency evaluation, the result for $\chi_{s/d}^{ME}$ does not resolve the two-peak structure.  This is not surprising, and exemplifies the ill-posed nature of numerical analytic continuation.  It remains an open question if improved numerical analytic continuation methods might resolve these distinctions.\cite{gull:nevanlinna, jiani:2021}  Our results throughout this manuscript do not suffer from this issue, since they represent true analytic continuation, the symbolic replacement of $i\omega_n \to \omega +i\Gamma$ where the choice of $\Gamma$ can in principle be made arbitrarily small. 

At lowest order the charge susceptibility represents a direct transition from an occupied state at some energy $\omega$ and momentum $k$ to an unoccupied state at $\omega+\Omega$ and momentum $k+q$.
At zero temperature the lowest energy where one will find an unoccupied state is the Fermi level.  At finite temperature there is of course a range of energies (on the scale of $T$) where unoccupied states will be available. 
Since the multi-peak structures in Figs.~\ref{fig:realfreq} and \ref{fig:realcut} occur at finite frequency they cannot be related to static Fermi surface nesting.  To elucidate their origin, we plot in Fig.~\ref{fig:spectralfunction} the spectral function for the case of $U/t=2.5$ at a density of $n=0.34$ where the observed splitting is strong.  Surprisingly, we observe that at the bottom of the band near $\mathbf{q}=(0,0)$ the dispersion becomes split, albeit weakly.  In this particular case, instead of a single quasi-particle peak at energy $\omega=\mu=-1.4 t$ we observe  two peaks, one at $\omega=-1.2t$ and another near $\omega \sim -1.8t$.  We can see that, in the case of the charge susceptibility, these peak locations correlate strongly with the peaks in Figs.~\ref{fig:realfreq} and \ref{fig:realcut}.  We surmise that finite energy nesting comes into play because the dispersion near $k=(0,0)$ and $(\pi,0)$ are both flat.  There is then a roughly fixed energy transition from each peak at the bottom of the band to the $(\pi,0)$ point - we illustrate this with dashed-red lines in Fig.~\ref{fig:spectralfunction}.  Further, the van Hove singularity occurs precisely when the dispersion at $\mathbf{q}=(\pi,0)$ meets the Fermi level and so this effect is expected to be strongest in the vicinity of the van Hove singularity of the Hartree-shifted starting point of the expansion.

\smallskip

\section{Conclusions}  We have presented a complete and consistent picture of the formation of an incompressible phase in single-particle properties of the $t-t^\prime-U$ Hubbard model on a 2D square lattice when in proximity to the van Hove singularity.  We have demonstrated the impact of the van Hove singularity on spin and charge excitations in the weak coupling limit of the 2D Hubbard model from a perturbative perspective. By considering static and dynamic properties of the model, we observe a disconnect between 
spin excitations that are strongest in proximity to half-filling from charge excitations that demonstrate a minima near the van Hove point.
Through a simple argument based on susceptibilities in the $\chi_{\uparrow\uparrow}$/$\chi_{\uparrow\downarrow}$ basis, it becomes clear that commensurate charge excitations should be expected on both the electron and hole doped sides of the phase diagram while spin excitations, at the temperatures explored, remain fixed at half-filling - although slightly on the hole-doped side for negative values of $t^\prime$.  We further examine the dynamical susceptibilities without the need for ill-posed numerical analytic continuation, providing access to plasmon and magnon dispersions.  We observe a splitting both of the plasmon and magnon dispersions into two distinct modes that merge in the $q\to 0 $ limit.  Such splitting is similar to what has been observed at stronger coupling strengths for charge excitations in 1D Hubbard chains due to the formation of a gap.\cite{Li:2021} 
We note that multiple peaks only occur for densities near or below the van Hove point.  Recent Raman experiments have suggested that the pseudogap phenomenon might terminate at the density associated with the Lifshitz point\cite{benhabib:2015} which for a weakly coupled system coincides with the van Hove singularity.  Although we are not directly computing the Raman spectra, our results would corroborate this observation.  
Our work has further implications to experimental probes of cuprates that have observed the existence of a high energy plasmon and low energy magnon for systems with particle-hole asymmetry.\cite{Lin:2020,lee:2014,Hepting2018}  Our results suggest that observing spin and charge excitations on the electron-doped side of the phase diagrams would require probing systems at higher energies than the hole-doped side.


\bibliographystyle{apsrev4-2}
\bibliography{refs.bib}

\end{document}